\documentclass[%
  reprint,
  superscriptaddress,
  twocolumn,
  amsmath,
  amssymb,
  10pt,
  aip,
  jcp,
  citeautoscript,
  notitlepage,
  longbibliography
]{revtex4-2}

\usepackage[utf8]{inputenc}
\usepackage[T1]{fontenc}
\usepackage{xcolor}
\usepackage[%
  colorlinks=true,
  allcolors=blue
]{hyperref}
\usepackage{url}
\usepackage{enumitem}
\usepackage{amsfonts}
\usepackage{amssymb}
\usepackage{amsmath}
\usepackage{mathtools}
\usepackage[ruled,vlined]{algorithm2e}
\usepackage{lmodern}
\usepackage[normalem]{ulem}

\newcommand{\rfig}[1]{Fig.~\ref{#1}}

\newcommand{\rref}[1]{Ref.~\citenum{#1}}

\newcommand{\bz}{\mathbf{z}}
\newcommand{\bx}{\mathbf{x}}

\newcommand{\chg}[1]{{\color{black}{#1}}}

\begin{document}
\title{%
  A Note on Spectral Map
}

% Authors, affiliations, addresses.
\author{Tuğçe Gökdemir}
\author{Jakub Rydzewski}
\email[Email: ]{jr@fizyka.umk.pl}
\affiliation{%
  Institute of Physics,
  Faculty of Physics, Astronomy and Informatics,
  Nicolaus Copernicus University,
  Grudziadzka 5, 87-100 Toru\'n, Poland
}

\date{\today}

\begin{abstract}
{\it Note:} A letter prepared for the \href{https://www.jstage.jst.go.jp/browse/mssj/\_pubinfo/-char/en}{Ensemble} journal of the Molecular Simulation Society of Japan (MSSJ).
\\[2pt]
In molecular dynamics (MD) simulations, transitions between states are often rare events due to energy barriers that exceed the thermal temperature. Because of their infrequent occurrence and the huge number of degrees of freedom in molecular systems, understanding the physical properties that drive rare events is immensely difficult. A common approach to this problem is to propose a collective variable (CV) that describes this process by a simplified representation. However, choosing CVs is not easy, as it often relies on physical intuition. Machine learning (ML) techniques provide a promising approach for effectively extracting optimal CVs from MD data. Here, we provide a note on a recent unsupervised ML method called spectral map, which constructs CVs by maximizing the timescale separation between slow and fast variables in the system. 
\end{abstract}

\maketitle

\section{Introduction}
Molecular dynamics (MD) allows us to observe properties of complex molecular systems at the atomistic level~\cite{dfrenkel:mc}. Over the recent years, many significant advances have been made in the development of advanced MD techniques~\cite{valsson2016enhancing} and specialized computing architectures~\cite{shaw2008anton}. Many processes investigated via MD, however, are characterized by infrequent transitions between their states as energy barriers often exceed thermal energy. To such processes, we can include conformational changes~\cite{lindorff2011fast} and nucleation~\cite{beyerle2023recent}. Therefore, despite the recent advancements, sampling such rare events presents a significant challenge due to the scarcity of observations and the huge number of degrees of freedom, which hampers our understanding. 

A common approach to this problem is to propose a collective variable (CV) that describes a process by a simplified representation. As the slow modes of the system mainly drive the dynamics, optimal CVs should be able to capture these processes~\cite{noe2017collective}. Extracting slow CVs is thus crucial for providing a useful and interpretable representation of the physical system. However, choosing CVs is difficult, as it often relies on physical intuition or trial and error. For instance, the radius of gyration and end-to-end distances are commonly selected as CVs during protein folding. Nevertheless, these CVs may not precisely capture the slow dynamics of the system, especially in complex molecular systems.

The selection of optimal CVs is crucial not only to simplify the representation of a complex molecular system but also to improve the accuracy and efficiency of sampling. As the timescale to overcome large energy barriers is significantly higher than the reachable timescale of MD simulations, a simulation can experience convergence issues~\cite{valsson2016enhancing}. To resolve these limitations, many enhanced sampling techniques have been proposed to capture rare events in a reasonable computational time~\cite{torrie1977nonphysical,voter1997hyperdynamics,bolhuis2002transition,earl2005parallel,barducci2008well,darve2008adaptive,allen2009forward}. Several of them, such as metadynamics~\cite{laio2002escaping}, umbrella sampling~\cite{mezei1987adaptive}, or variationally enhanced sampling~\cite{valsson2014variational} require a few CVs before the simulation and thus we refer to them as CV-based enhanced sampling techniques.

Machine learning (ML) offers a potential solution by effectively reducing the dimensionality of complex physical systems. Recently, there has been significant research interest in learning CVs directly from the simulation data of high-dimensional physical systems. Ongoing developments in these methods continue to advance this field~\cite{rydzewski2023manifold,mehdi2024enhanced}. Several methods use temporal information, such as time-series data, to construct slow CVs. These methods include time-lagged independent component analysis~\cite{naritomi2011slow,hernandez2013identification}, time-lagged autoencoders~\cite{wehmeyer2018time}, VAMPnets~\cite{mardt2018vampnets}, and state-predictive information bottleneck~\cite{wang2021state}. Another class of methods that do not require temporal data includes techniques such as Laplacian eigenmap~\cite{belkin2003laplacian,belkin2001laplacian}, diffusion map~\cite{coifman2005geometric,coifman2006diffusion,nadler2006diffusion}, and reweighted stochastic embedding~\cite{zhang2018unfolding,rydzewski2021multiscale,rydzewski2022reweighted,rydzewski2023selecting}. Additionally, we can include a recently developed technique called spectral map (SM)~\cite{rydzewski2023spectral,rydzewski2024learning,rydzewski2024spectral}, which is the focus of this note.

\section{Spectral Map}

\begin{figure*}
  \centering
  \includegraphics[width=\textwidth]{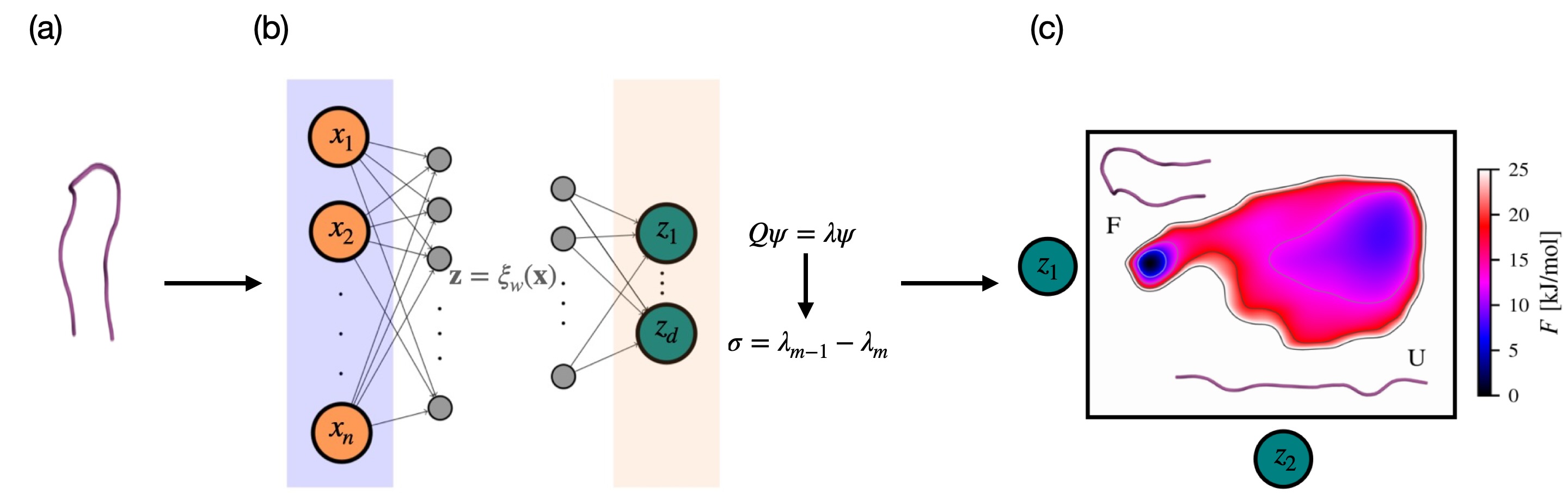}
  \caption{Example of SM learning illustrated on the mini-protein chignolin dataset. (a) Input for SM consisting of conformations from the MD trajectory. (b) High-dimensional data is reduced to the low-dimensional space via the shown NN. The spectral gap $\sigma$ is calculated from the dominant eigenvalues of the transition matrix. The NN is trained iteratively by maximizing the spectral gap and backpropagation. (c) The corresponding free-energy landscape of chignolin shows two distinct free-energy basins: the folded state (F) and the unfolded state (U). The figure is based on results presented in \rref{rydzewski2024learning}.}
  \label{fig}
\end{figure*}

In SM, a high-dimensional system described by $n$ configuration variables $\bx=(x_1, \dots, x_n)$ is mapped to $d$ CVs $\bz=(z_1, \dots, z_d)$ by a feed-forward neural network (NN): 
\begin{equation}
\mathbf{z} = \xi_w(\mathbf{x}) \equiv \{\xi_k(\mathbf{x}, w)\}_{k=1}^d,
\end{equation}
where $w$ are trainable parameters. To estimate the extent of the timescale separation between slow and fast variables, SM constructs a Gaussian kernel in CV space $g(\mathbf{z}_k, \mathbf{z}_l) = \exp\left(-{\|\mathbf{z}_k - \mathbf{z}_l\|^2/\varepsilon}\right)$, where $\varepsilon$ is a scale constant. Then, the anisotropic diffusion kernel is used to enocde data density into CV space:
\begin{equation}
  \kappa(\mathbf{z}_k,\mathbf{z}_l) = \frac{g(\mathbf{z}_k,\mathbf{z}_l)}{\sqrt{\rho(\mathbf{z}_k)\rho(\mathbf{z}_l)}}, 
\end{equation}
where $\rho(\bz_k) = \sum_l g(\bz_k, \bz_l)$ is a kernel density estimate. Finally, a transition matrix is constructed by the row normalization of the anisotropic diffusion kernel:
\begin{equation}
\label{eq:reducedmarkov}
  Q(\mathbf{z}_k,\mathbf{z}_l) = \frac{\kappa(\mathbf{z}_k,\mathbf{z}_l)}{\sum_i \kappa(\mathbf{z}_k,\mathbf{z}_i)},
\end{equation}
where each entry denotes the transition probability $\mathrm{Pr}\left( \mathbf{z}_{i+1} = \mathbf{z}_l \mid \mathbf{z}_i = \mathbf{z}_k \right)$ from $\mathbf{z}_k$ to $\mathbf{z}_l$ in the low-dimensional data.

As the scoring function to maximize during training, SM uses the spectral gap, which is estimated as the difference between neighboring eigenvalues of the transition matrix:
\begin{equation}
  \label{eq:sg}
\sigma = \lambda_{m-1} - \lambda_{m},
\end{equation}
where the eigenvalues are sorted in descending order $\lambda_0 = 1 \geq \lambda_1 \geq \lambda_2 \geq \ldots$ and $m$ denotes the number of metastable states in reduced space. Maximizing the spectral gap allows us to learn iteratively improved slow CVs. The training algorithm is shown in Algorithm~\ref{alg:spectral-map} with an illustration presented in \rfig{fig}, showing slow CVs constructed from a dataset provided by DE Shaw Research~\cite{lindorff2011fast}. A reference implementation of SM is available on PLUMED-NEST (\url{www.plumed-nest.org}), the public repository of the PLUMED consortium, as plumID:24.005~\cite{plumed,plumed-nest}.

\begin{algorithm}
  \SetKwInOut{Input}{Input}
  \SetKwInOut{Output}{Output}
  \Input{\chg{Dataset, number of CVs $d$, number of metastable states $m$.}}
  \Output{NN $\xi_{w}(\bx)=\bz$ for CVs with maximal spectral gap.}
  \begin{minipage}{0.9\linewidth}
  \begin{enumerate}
    \item Iterate over training epochs:
      \begin{enumerate}
        \item Map samples to their reduced representation using the NN.
        \item Iterate over data batches:
      \begin{enumerate}
        \item Calculate the Markov transition matrix $Q$ from the anisotropic diffusion kernel in CV space.
        \item Perform eigendecomposition and estimate the spectral gap as $\sigma=\lambda_{m-1} - \lambda_m$.
        \item Update weights $w$ by maximizing the score given by the spectral gap.
      \end{enumerate}
    \end{enumerate}
    \item Return the NN if the learning converged.
  \end{enumerate}
  \end{minipage}
  \caption{Spectral map learning algorithm}
  \label{alg:spectral-map}
\end{algorithm}

As in diffusion maps, SM uses the anisotropic diffusion kernel, but it is calculated in CV space. Diffusion maps approximate CV space by using eigenfunctions computed in configuration space. Since configuration space remains constant, the reliability of the eigenfunctions cannot be verified self-consistently, and their quality cannot be refined. Moreover, SM does not use eigenfunctions to parametrize CV space. Rather, SM is similar to parametric dimensionality reduction techniques such as reweighted stochastic embedding~\cite{zhang2018unfolding,rydzewski2021multiscale,rydzewski2022reweighted,rydzewski2023selecting} that use NNs to construct CVs. As CVs calculated via SM are represented by an NN, they can be readily used in CV-based enhanced sampling techniques. It is worth noting that spectral gap optimization of order parameters (SGOOP) introduced by Tiwary and Berne~\cite{tiwary2016spectral} also maximizes the timescale separation to find slow CVs using a spectral gap estimate similar to that of SM.

\section{Conclusion}
In this note, we have briefly recapped SM, an unsupervised deep ML framework to construct CVs by maximizing the timescale separation between slow and fast variables in the system. More detailed discussion is available in works by Gökdemir and Rydzewski~\cite{rydzewski2023spectral,rydzewski2024learning,rydzewski2024spectral}. We think that SM is an interesting approach to learning slow CVs and can be valuable for the analysis of molecular systems. Currently, our work focuses on using SM on-the-fly during MD simulations to improve the accuracy and efficiency of enhanced sampling aimed at discovering long-timescale processes.

\section*{Acknowledgment}
The research was supported by the National Science Center in Poland (Sonata 2021/43/D/ST4/00920, ``Statistical Learning of Slow Collective Variables from Atomistic Simulations''). J. R. acknowledges funding the Ministry of Science and Higher Education in Poland. The authors thank Atsushi Ito for the interesting discussions and the invitation to publish this note in the Ensemble journal of the Molecular Simulation Society of Japan (MSSJ).

\bibliography{main.bib}

\end{document}